# On infrared pseudogap in cuprate and pnictide high-temperature superconductors

S. J. Moon,[1,2] Y. S. Lee,[1,3] A. A. Schafgans,[1] A. V. Chubukov,[4] S. Kasahara,[5] T. Shibauchi,[6] T. Terashima,[5] Y. Matsuda,[6] M. A. Tanatar,[7] R. Prozorov,[7] A. Thaler,[7] P. C. Canfield,[7] S. L. Bud'ko,[7] A. S. Sefat,[8] D. Mandrus,[8,9] K. Segawa,[10] Y. Ando,[10] and D. N. Basov[1]

[1]*Department of Physics, University of California, San Diego, La Jolla, California 92093, USA*

[2]*Department of Physics, Hanyang University, Seoul 133-791, South Korea*

[3]*Department of Physics, Soongsil University, Seoul 156-743, Republic of Korea*

[4]*Department of Physics, University of Wisconsin-Madison, 1150 Univ. Ave., Madison, Wisconsin 53706-1390, USA*

[5]*Research Center for Low Temperature and Materials Science, Kyoto University, Kyoto 606-8502, Japan*

[6]*Department of Physics, Kyoto University, Kyoto 606-8502, Japan*

[7]*Ames Laboratory and Department of Physics and Astronomy, Iowa State University, Ames, Iowa 50011, USA*

[8]*Materials Science and Technology Division, Oak Ridge National Laboratory, Oak Ridge, Tennessee 37831, USA*

[9]*Department of Materials Science and Engineering, University of Tennessee, Knoxville, Tennessee 37996, USA*

[10]*Institute of Scientific and Industrial Research, Osaka University, Ibaraki, Osaka 560-0047, Japan*

We investigate infrared manifestations of the pseudogap in the prototypical cuprate and pnictide superconductors: YBa$_2$Cu$_3$O$_y$ and BaFe$_2$As$_2$ (Ba122) systems. We find remarkable similarities between the spectroscopic features attributable to the pseudogap in these two classes of superconductors. The hallmarks of the pseudogap state in both systems include a weak absorption feature at about 500 cm$^{-1}$ followed by a featureless continuum between 500 and 1500 cm$^{-1}$ in the conductivity data and a significant suppression in the scattering rate below 700 – 900 cm$^{-1}$. The latter result allows us to identify the energy scale associated with the pseudogap $\Delta_{PG}$. We find that in the Ba122-based materials the superconductivity-induced changes of the infrared spectra occur in the frequency region below 100 – 200 cm$^{-1}$, which is much lower than the energy scale of the pseudogap. We performed theoretical analysis of the scattering rate data of the two compounds using the same model which accounts for the effects of the pseudogap and electron-boson coupling. We find that the scattering rate suppression in Ba122-based compounds below $\Delta_{PG}$ is solely due to the pseudogap formation whereas the impact of the electron-boson coupling effects is limited to lower frequencies. The magnetic resonance modes used as inputs in our modeling are found to evolve with the development of the pseudogap, suggesting an intimate correlation between the pseudogap and magnetism.

## I. INTRODUCTION

Since the discovery of superconductivity in the Cu oxides (cuprates), the physics of high-transition-temperature ($T_c$) superconductivity has been a major theme in the modern condensed matter physics.[1-4] In spite of tremendous progress over the nearly past three decades, unconventional normal and superconducting properties observed in cuprates are yet to be fully understood. In 2008, a new class of high-$T_c$ superconductors, iron pnictides/chalcogenides has been discovered.[5] The analysis of common and contrasting trends of these two classes of superconductors is essential as it may help to identify the primal aspects of high-$T_c$ superconductivity.[6]

One of the enigmatic properties in the cuprate is pseudogap state occupying a substantial portion of the cuprate phase diagram.[7-13] While an unprecedentedly large number of experiments have addressed the existence of the pseudogap phase and relevant physical phenomena, the nature of the pseudogap and its relation to high-$T_c$ superconductivity remain unresolved.[7,12] Despite its yet undetermined origin, the pseudogap is universally regarded as an essential piece of the physics of unconventional cuprate superconductors. There is mounting evidence that superconductivity in iron pnictides/chalcogenides is also different from a conventional *s*-wave.[14,15] Recently, we reported on infrared pseudogap in the response of a prototypical pnictide high-$T_c$ superconductor: Ba122 systems.[16] In Ref. 16, we discussed the infrared characteristics of the pseudogap in this family of Fe-based superconductors (Fe-SC) and the cuprates on the same footing driven by the goal to explore similarities and differences in the manifestation of the pseudogap in these two classes of high-$T_c$ materials. Empowered with this analysis we attempted to narrow down the field of possible microscopic scenarios of the mysterious pseudogap phase.

High-$T_c$ superconductivity in both the cuprates and the iron pnictides emerges when the antiferromagnetic (AFM) order of parent compounds is suppressed. However, the electronic states of the respective parent phases are drastically different: Mott insulator in the cuprates and bad metals in the pnictides.[8, 14, 15] In the cuprates, the pseudogap has been attributed to the persistence of the parent compound gap in doped phases,[1] the precursor of the superconducting gap in the normal state,[17] the precursor to antiferromagnetism,[18-20] circulating loop currents,[21] and charge-density wave order (or quasi-order) which competes with superconductivity.[22, 23] The evidence for a competing charge-density-wave order has emerged in recent years,[24-27] and several theoretical scenarios for the competition between charge order and superconductivity have been put forward.[28-34]

Fundamentally, doped cuprates are correlated metals with strongly renormalized electronic bands.[35] Normal state properties of the pnictides are also dominated by correlations.[6, 36] Furthermore, the phase diagram of the Fe-SCs reveals a number of common attributes with the cuprates. The bad-metal[17] parent compounds of the Fe-SCs show a gap feature that is associated with an AFM spin-density-wave order.[37-41] With charge carrier or isovalent doping into the mother compounds, the spin-density-wave order in Fe-based systems is gradually suppressed and the superconductivity emerges. It is believed that the magnetic fluctuations play a role in the superconducting pairing mechanism by mediating the effective interaction between fermions that leads to superconductivity.[42, 43] Based on the results of infrared studies of the Ba122 systems, we suggested that the AFM precursor, i.e., short-range AFM instability, may be responsible for the pseudogap in the Fe-SCs.[16] We will critically examine and elaborate on this conjecture while exploring impact of various forms of energy (pseudo)gaps in infrared observables collected for several different classes of correlated electron systems (Fig. 4).

The pseudogap in the infrared response of the cuprates has been well documented, and advanced analysis needed to unveil pseudogap features in the raw data has been developed.[7, 11, 44-46] On the other hand, there have been only few infrared studies of the pseudogap in the Fe-SCs. In Ref. 16, we identified the low-energy (< 60 meV) infrared pseudogap feature in the Ba122 system by employing the analyses previously used for the cuprates. This finding allows us to make the direct comparison between the infrared signatures of the pseudogap formation in the two classes of high-$T_c$ superconductors, aiming at obtaining insights into the mechanism of the pseudogap phase. This low-energy infrared pseuodgap is distinct from higher frequency ($\omega$>150-200 meV) characteristics of the pnictides apparent through the analysis of the redistribution of the electronic spectral weight (see Section VI below).

In this article, we analyze on the same footing the infrared spectra of $YBa_2Cu_3O_y$ ($YBCO_y$) and Ba122 systems, and find surprising similarities between the two data sets. In both compounds, the pseudogap formation in the normal state leads to a weak absorption feature in the conductivity data and a distinct suppression in the scattering rate. Investigations of the electrodynamics above and below $T_c$ reveal that the energy scale of the infrared pseudogap is much higher than that of the superconducting gap in both the systems. The theoretical analysis accounting for the doping dependence of the infrared pseudogap and electron-boson coupling further reveals a correlation between the pseudogap and magnetism.

The paper starts with the description of experimental details in Sec. II. We review overall trends in the evolution of the optical conductivity with doping in Sec. III. The infrared spectroscopic data and analyses associated with the pseudogap formation are presented in Sec. IV, followed by discussion on the relationship between the infrared pseudogap and superconductivity in Sec. V. A critical survey of the pseudogap characteristics found in our infrared data and various other probes is provided in Sec. VI. Nature of the pseudogap phase is discussed in Sec. VII. Summary is presented in Sec. VIII.

## II. EXPERIMENTAL

YBCO$_y$ single crystals with oxygen content $y = 6.28 - 7.00$ were grown by a conventional flux method in Y$_2$O$_3$ crucibles and detwinned under uniaxial pressure. Details of growth procedures and characterization are described elsewhere.[47] Detwinned single crystals allow us to investigate the response of the CuO$_2$ plane (*a* axis) unperturbed by contributions due to the chains or chain fragments along the *b* axis. In this paper, we focus on the *a*-axis optical spectra. High-quality single crystals of BaFe$_2$(As$_{0.67}$P$_{0.33}$)$_2$ (P-Ba122) and Ba(Fe$_{1-x}$Co$_x$)$_2$As$_2$ (Co-Ba122) were grown using self-flux method. Details of growth procedures and characterization results are described in Refs. 48-51. Parent compound Ba122 ($x=0$) exhibits structural and magnetic transition at $T_N=135$ K.[50] Optimally doped (OPD) Co-Ba122 ($x=0.08$) and P-Ba122 show superconducting transition at $T_c=22$ and 30 K, respectively.[48, 49] Overdoped (OD) Co-Ba122 ($x=0.25$) is non-superconducting.[52] The in-plane reflectance $R(\omega)$ spectra of these crystals were measured at various temperatures (*T*) using in-situ overcoating technique.[53] The complex optical conductivity $\sigma_1(\omega)+i\sigma_2(\omega)$ was determined from the Kramers-Kronig analysis of $R(\omega)$.[54] The reflectance and conductivity spectra of YBCO$_y$ were originally reported in Ref. 55. The reflectance spectra of the Ba122 compounds have not been published before. The theoretical analysis using the formalism devised in Ref. 56 is new and enables direct comparison between YBCO$_y$ and the Ba122 systems.

### III. EVOLUTION OF OPTICAL SPECTRA WITH DOPING

We start our discussion by describing the doping-induced changes in raw reflectance spectra $R(\omega)$ of the two classes of high-$T_c$ superconductors. The room-temperature $R(\omega)$ data of YBCO$_y$ compounds are displayed in Fig. 1(a). The parent compound of YBCO$_y$ system is a Mott insulator. In the heavily underdoped YBCO$_{6.28}$, the absolute reflectance level is low and sharp phonon resonances are identified in the far-infrared response. The far-infrared reflectance rises toward zero frequency, indicating a weakly metallic response. In the mid/near-infrared frequency region, $R(\omega)$ decreases gradually at frequencies below the onset of charge transfer excitation near 12000 cm$^{-1}$ (1.5 eV). Upon charge carrier doping, $R(\omega)$ data show drastic changes with the elevation of the level of $R(\omega)$ and the contribution of the phonon resonances becomes less pronounced. Concurrently, the reflectance minimum, so called plasma minimum near 10000 cm$^{-1}$ shifts to higher frequencies.

The reflectance spectra of Ba122 system shown in Fig. 1(b) reveal much weaker doping dependences than those of YBCO$_y$. The reflectance of the undoped Ba122 shows metal-like behavior, which is consistent with other infrared data.[38-41, 57] Electron doping (Co substitution) or isovalent doping (P substitution) leads to an increase in $R(\omega)$ below about 2000 cm$^{-1}$, suggesting the enhancement of coherent electronic response.

Figure 2(a) shows the real part of the room-temperature conductivity spectra of YBCO$_y$ calculated from Kramers-Kronig analysis of $R(\omega)$. The sharp peaks below 700 cm$^{-1}$ are due to transverse optical phonons. In $\sigma_1(\omega)$ of YBCO$_{6.28}$, the charge transfer gap with the magnitude of about 2 eV (~ 16000 cm$^{-1}$) is clearly observed (Fig. 2(a)). Even at the weakest doping levels, a Drude-like mode is apparent in $\sigma_1(\omega)$ spectra. The corresponding ground state is referred to as nodal metal[55] since nodal quasiparticles dominate in the transport and spectroscopic data for weakly doped YBCO$_y$. We also observe a localized absorption mode in the mid-infrared region near 0.5-0.6 eV (4000-5000 cm$^{-1}$) in the $\sigma_1(\omega)$ spectra for weakly doped phases. Upon further doping, the mid-infrared modes acquire an increasing amount of spectral weight and show a significant softening. With doping increasing above $y$~6.70, the mid-infrared modes merge into a broad background and separate modes are impossible to identify. Near optimal doping, the featureless mid-infrared conductivity is found, which is a common property of electromagnetic response of most cuprates in this doping regime.[7]

Room-temperature conductivity spectra of Ba122 compounds are displayed in the bottom panel of Fig. 2. In the entire doping regime from parent compound to non-superconducting overdoped material, the far-infrared response is dominated by a relatively narrow Drude-like feature followed by a smooth mid-infrared continuum in the frequency region between 700 and 1500 cm$^{-1}$. A strong absorption centered at about 5000 cm$^{-1}$ can be assigned as an interband transition from As $p$ to Fe $d$ states.[58] The doping leads to mild enhancement of the Drude-like response and weak suppression of the mid-infrared continuum.

A notable similarity found in our $\sigma_1(\omega)$ data of YBCO$_y$ and Ba122 compounds is the coexistence of the coherent mode centered at $\omega$=0 and featureless mid-infrared conductivity. Similar response has been observed in various classes of correlated electron systems.[35] We stress that both in the cuprates and the Fe-SCs the coherent mode and incoherent background are the two constituents of intraband conductivity.[16, 37, 59] The coherent component is represented by the Drude-like mode. The remainder of the itinerant-carrier spectral weight is spread out over broad frequency region in mid-infrared frequencies. It is well known that the cuprates are single-band system where $x^2$-$y^2$ band solely contributes to the electrodynamics.[7] Whereas the Fe-SCs are multiband systems,[60] spectral weight analyses of the Ba122 systems showed that the contribution of itinerant carriers extends to mid-infrared region in the form of featureless, incoherent conductivity.[16, 37, 61]

The spectral weight associated with the itinerant carrier response can be used to quantify the impact of many-body effects.[36] It is instructive to inquire into the value of $K_{exp}/K_{LDA}$, where $K_{exp}$ and $K_{LDA}$ are experimental and non-interacting theoretical kinetic energies, respectively. The ratio of $K_{exp}/K_{LDA}$ allows one to quantify narrowing of the electronic bands due to correlation effects.[35, 62] The magnitude of $K_{exp}$ is obtained from the integration of $\sigma_1(\omega)$ up to a cutoff frequency chosen to accommodate the entire intraband spectral weight. In Ba122 materials the degree of the kinetic energy renormalization can be consistent with other experimental results only by attributing the spectral weight at $\omega \lesssim 1500$ cm$^{-1}$ to the itinerant carrier response.[37] The kinetic energy analyses of the cuprates and the iron pnictides reveal interesting common trends: for materials with highest $T_c$ in each family, the value of $K_{exp}/K_{LDA}$ is found to be $0.3 - 0.5$.[6] This value implies that optimal degree of correlations in between purely localized and purely itinerant regimes may be essential for high-$T_c$ superconductivity.[36]

## IV. MANIFESTATIONS OF PSEUDOGAP IN OPTICAL SPECTRA

### A. Optical conductivity

Following N. F. Mott, the term "pseudogap" implies a partial gap in the electronic density of states (DOS) and can be directly probed by spectroscopic techniques.[9, 63] However, in the high-$T_c$ literature the term "pseudogap" is commonly used in different contexts. As far as spectroscopic studies are involved it is customary to discriminate between the so-called low-energy (40-70 meV) and high-energy (100-150 meV) pseuodgaps. In this section, we examine the manifestations of the low-energy pseudogap in the infrared responses of YBCO$_y$ and Ba122 systems. We will elaborate on high-energy pseudogap in Section VI.

The first spectroscopic evidence for the formation of the electronic pseudogap was reported by Homes *et al.* and was obtained through the analysis of the *c*-axis infrared data of YBCO$_{6.70}$.[64] Due to the strictly incoherent nature of the *c*-axis transport of the underdoped cuprates, the conductivity data bear a more direct connection to the electronic DOS. The *c*-axis conductivity showed significant depression below the characteristic temperature $T^*$ and the appearance of the conductivity onset in the *c*-axis conductivity of YBCO$_{6.70}$ in the pseudogap phase. On the other hand, in case of the *ab*-plane response, the manifestation of the pseudogap formation is rather subtle.[11, 65, 66] Angle-resolved photoemission spectroscopy (ARPES) measurements reported the gap opening in the anti-nodal region far above $T_c$ while nodal region remains gapless.[67] Even though infrared experiments probe the momentum-space averaged response, the indications of the pseudogap can still be captured in the *ab*-plane infrared data[11, 59, 65, 66] as we will demonstrate below.

In order to highlight signatures of the infrared pseudogap in the *ab*-plane response, we focus on the *T* dependence of the in-plane conductivity $\sigma_1(\omega)$ of underdoped YBCO$_{6.65}$ and OPD P-Ba122. Figure 3(a) displays $\sigma_1(\omega)$ of YBCO$_{6.65}$ of which pseudogap temperature $T^*$ is about 300 K. As *T* decreases, the coherent Drude-like response narrows leading to a rapid increase of the DC conductivity. At *T*=65 K, we observe a weak absorption feature in the mid-infrared continuum of the conductivity: a slight upturn in $\sigma_1(\omega)$ at 500 cm$^{-1}$ followed by a plateau. The same trend is registered in $\sigma_1(\omega)$ spectra for P-Ba122.[16] The absorption feature indicated by an arrow in Fig. 3(b) is clearly resolved in the conductivity data at *T*≤100 K. We attribute the weak absorption and featureless mid-infrared conductivity to the formation of the infrared pseudogap.

The spectroscopic characteristics of the low-energy pseudogap in the response of the cuprates are quite distinct from those of various other gaps observed in condensed matter systems. Figure 4 shows the manifestations of the optical gap for several instructive cases including superconducting gap, density-wave gap, hybridization gap, Mott gap, and Pauli blocking gap. The changes in $\sigma_1(\omega)$ and schematics of DOS are shown in the top and bottom panels, respectively. The onset of superconductivity in dirty *s*-wave superconductors leads to the drastic suppression of the conductivity below the superconducting gap 2Δ$_{SC}$ followed by an abrupt threshold structure (Fig. 4(b)). The missing spectral weight forms superfluid $\delta$-peak at zero frequency[7]. For the cases of the spin-density-wave (SDW) gap in Cr and the hybridization gap in YbFe$_4$Sb$_{12}$,[45, 68] one can see the significant suppression in the optical conductivity below the gap and the appearance of a narrow Drude-like peak. The narrowing of the Drude-like peak can be attributed to the decrease in the amount of states available for scattering of remaining quasiparticles due to the formation of a partial gap. The spectral weight piles up right above the gap leading to conspicuous absorption peaks (Figs. 4(c) and 4(d)). Figure 4(e) shows the changes in $\sigma_1(\omega)$ due to bandwidth-controlled Mott transition in $\kappa$-(BEDT-TTF)$_2$Cu[N(CN)$_2$]Br$_{1-x}$Cl$_x$.[69, 70] In the most metallic compound ($x$=0.9), coherent Drude-like and flat incoherent responses are present in the far- and mid-infrared regions, respectively. The mid-infrared conductivity originates from optical transitions between lower and upper Hubbard bands. As the electronic state of the system changes from metal to insulator, the Drude-like response becomes suppressed and the optical transition between the Hubbard bands is enhanced. In transition metal oxides with 3$d$ orbitals, the on-site Coulomb repulsion $U$ is strong and the spectral weight transfer between Drude-like mode and Mott-Hubbard excitations occurs over a broad energy range comparable to $U$. Optical spectroscopic experiments on several prominent 3$d$ transition metal oxides, such as VO$_2$, V$_2$O$_3$, and $R$NiO$_3$ ($R$: rare earth elements) indeed showed that the energy scale is as large as 5 – 6 eV.[71-73] In the case of graphene, the conductivity data show gap-like threshold structure by electrostatic or chemical doping although the DOS remains gapless (Fig. 4(f)). When the Fermi level is at the charge neutral point, the conductivity is frequency-independent due to linear dispersion of the top and bottom cones. When the Fermi level is shifted by external bias, a threshold structure appears.[74] This form of the conductivity is consistent with the notion of Pauli blocking: direct interband transitions between the bottom and top cones are forbidden by momentum conservation for $\hbar\omega < 2E_F$. The spectral weight lost from $\hbar\omega < 2E_F$ is transferred to Drude-like peak due to mobile Dirac quasiparticles; steep increase in the imaginary conductivity $\sigma_2(\omega)$ at low frequencies indicates the Drude-like absorption (Fig. 4(l)).

Note that in all cases displayed in Fig.4 the complex conductivity data $\sigma_1(\omega)+i\sigma_2(\omega)$ allow one to unambiguously determine the destination of the spectral weight "missing" from the intra-gap region. In some cases, this missing weight can be readily identified by examining the spectra of the real part of the conductivity. However, this is not always possible since the "missing" spectral weight can be recovered outside of experimentally accessible region in the spectra of $\sigma_1(\omega)$. If latter is the case then it is instructive to examine the imaginary part of the conductivity. Provided the missing spectral weight is recovered below the low-energy cut-off of the $\sigma_1(\omega)$ data, the imaginary part of the conductivity reveals a characteristic increase that can be recognized in the frequency range where actual data exist. This is most clearly exemplified in Fig.4 for the superconducting gap and the Pauli blocking gap.

We stress that the formation of the infrared pseudogap in YBCO$_y$ and Ba122 systems does not exhibit clear signatures of spectral weight transfer in $\sigma_1(\omega)$. This is in contrast with the formation of the other gaps in Fig. 4, where the spectral weight transfer over an energy region comparable to the gap energy is noticeable and also leads to the appearance of pronounced absorption peak.

### B. Scattering rate

In many materials the signatures of the low-energy infrared pseudogap are best resolved in the frequency-dependent scattering rate $1/\tau(\omega)$. One can obtain $1/\tau(\omega)$ using the extended Drude model (EDM),

$$\frac{1}{\tau(\omega)} = \frac{\omega_p^2}{4\pi} \mathrm{Re}\left(\frac{1}{\sigma_1(\omega)+i\sigma_2(\omega)}\right), \tag{1}$$

where $\omega_p$ is the plasma frequency and $\tilde{\sigma}(\omega)$ is the complex optical conductivity.[11] For this analysis, the value of $\omega_p$ is estimated by integrating the $\sigma_1(\omega)$ up to a certain frequency (8000

cm$^{-1}$ for YBCO and 1500 cm$^{-1}$ for P-Ba112). The cutoff frequencies are chosen to account only the intraband contribution to $\sigma_1(\omega)$ of each compound.[16, 55] We modeled interband transitions using Lorentz oscillators and subtracted their contribution from the conductivity data of the Ba122 compounds as described in Ref. 16. A recent optical study of Co-Ba122 compounds suggested that interband transition may exist at around 1000 cm$^{-1}$.[75] As pointed in Ref. 75, the estimation of the contribution from the interband transition at 1000 cm$^{-1}$ to optical conductivity has considerable uncertainties. Accurate estimation of the contribution may lead to better agreement between the experimental and theoretical scattering rate shown in Fig. 3. It should be pointed that the extended Drude model is used to analyze the optical response of systems where intranband response originates from a single band. Due to possible spurious effects in the scattering rate, the application of the extended Drude model to multiband systems should be cautious. Several infrared studies of the pnictides employed the multi-component analysis of the conductivity data.[39, 40, 75, 76] However, we believe that the single-band analysis is applicable to the Ba122 pnictides and reveals important physics.[16, 61, 77]

The pseudogap with the magnitude $\Delta_{PG}$ results in a suppression of $1/\tau(\omega)$ at low frequencies, leading to the appearance of a distinct threshold structure; the frequency of this threshold quantifies the energy scale of the pseudogap $\Delta_{PG}$.[11] This characteristic behavior is reproduced in the scattering rate spectra of both YBCO$_{6.65}$ and P-Ba122 as shown in Figs. 3(c) and 3(d). With decreasing $T$, $1/\tau(\omega)$ is depressed below about 900 cm$^{-1}$ and 700 cm$^{-1}$ for YBCO$_{6.50}$ and P-Ba112, respectively, revealing the threshold structures. The characteristic energy scales associated with these structures are indicated by gray bars.

We stress that the manifestations of the infrared pseudogap in $1/\tau(\omega)$ spectra are markedly different from those of the other gaps, as shown in the middle panels of Fig. 4. The opening of the superconducting, density-wave, hybridization, and Mott gap leads to the depression of $1/\tau(\omega)$ in the intragap region, followed by an overshoot of the spectra right above the gaps. This latter feature most likely reflects sharp features of the DOS.[45, 46] In contrast, the overshoot of $1/\tau(\omega)$ is not observed in the pseudogap case. This latter finding is consistent with notion that the formation of the pseudogap only involves gradual depression of the electronic DOS and is not accompanied by anomalies characteristic to superconducting or density wave gaps.

### C. Pseudogap and electron-boson coupling effects

The spectral form of the scattering rate is affected by coupling to resonant excitations as well as by the pseudogap.[11, 78-81] For example, a sharp mode in the bosonic spectral function $\alpha^2 F(\omega)$ induces a rapid increase in the scattering rate at frequencies above this mode. A formalism taking into account the combined effect on $1/\tau(\omega)$ from $\alpha^2 F(\omega)$ and from the modification of the density of states by the pseudogap has been developed by Sharapov and Carbotte.[56] They found:

$$\frac{1}{\tau(\omega,T)} = \frac{\pi}{\omega}\int_0^\infty d\Omega \alpha^2 F(\Omega) \int_{-\infty}^\infty d\varepsilon [N(\varepsilon-\Omega)+N(-\varepsilon+\Omega)]$$

$$\times [n_B(\Omega)+n_F(\Omega-\varepsilon)][n_F(\varepsilon-\omega)-n_F(\varepsilon+\omega)], \quad (2)$$

where $N(\varepsilon)$ is the normalized DOS, $n_B(\varepsilon)=1/(e^{\varepsilon/kT}-1)$ and $n_F(\varepsilon)=1/(e^{\varepsilon/kT}+1)$ are boson and fermion occupation numbers, respectively.

We apply Eq. (2) to model the experimental $1/\tau(\omega)$ spectra of YBCO$_y$ and Ba122 systems. We obtained the most accurate fit using the following form for $\alpha^2 F(\omega)$: a sum of a Gaussian peak and a background,

$$\alpha^2 F(\omega) = \frac{A}{\sqrt{2\pi}(d/2.35)} e^{-(\omega-\omega_1)^2/[2(d/2.35)^2]} + \frac{I_s\omega}{\omega^2+\omega_2^2}, \quad (3)$$

In Eq. (3), $A$ is the area under the Gaussian peak, $d$ is the full width at half maximum, and $\omega_1$ is the frequency position of the Gaussian resonance. $I_s$ and $\omega_2$ are the intensity and the characteristic frequency of the background spectrum, respectively.[82] This spectral form of $\alpha^2 F(\omega)$ is adopted from neutron studies of magnetic excitations of each compound.[83-86] We note that the neutron scattering data of the pnictides revealed a resonance mode at the AFM wavevector of the parent compound[83-85, 87] and a broad background.[83, 88-90] In Ref. 16, we attempted to fit the data for Ba122 pnictides including only the resonance mode. Here, we have chosen to include a broad background term as well, in close similarity with the cuprates. We find that the new model of $\alpha^2 F(\omega)$ spectra improves the agreement between the experimental and calculated $1/\tau(\omega)$ spectra of OPD P-Ba122. It should be mentioned that whereas the resonance mode is the most evident at $T<T_c$, it is still present in the normal state of YBCO$_y$ and Co-Ba122.[83, 91] In order to describe the impact of the pseudogap on the DOS, we used a quadratic gap function,[82]

$$N(\varepsilon) = \left[ N(0) + \left( (1-N(0))\frac{\varepsilon^2}{(\Delta_{PG}/2)^2} \right) \right] \theta(\frac{\Delta_{PG}}{2} - |\varepsilon|) + \theta(|\varepsilon| - \frac{\Delta_{PG}}{2}), \qquad (4)$$

where $\theta(\varepsilon)$ is the Heaviside function.

Figures 3(e) and 3(f) show the results of the analysis using Eqs. (2)-(4) on YBCO$_{6.65}$ and P-Ba122, respectively. The spectral forms of $\alpha^2 F(\omega)$ and DOS employed in the analysis are displayed in Figs. 3(g) and 3(h). In this analysis, we used $\Delta_{PG}=700$ cm$^{-1}$ for both compounds. The effect of the coupling to a sharp mode in $\alpha^2 F(\omega)$ is a steep onset of $1/\tau(\omega)$ at low frequencies. The broad background is responsible for a more gradual onset in $1/\tau(\omega)$ at low frequencies as well as a mild increase at higher frequencies. The main consequence of the pseudogap is to suppress $1/\tau(\omega)$ at frequencies below $\Delta_{PG}$. The relative roles of the bosonic coupling and of the pseudogap formation are well demonstrated in the calculated $1/\tau(\omega)$ with $\Delta_{PG}$=0. In the absence of the pseudogap, $1/\tau(\omega)$ increases rapidly at the low frequencies where the sharp mode is present. This behavior is distinct from the gradual increase in $1/\tau(\omega)$ over broader frequency range. The analyses demonstrate that three ingredients: the pseudogap, a resonance mode and a broad background in the spectral function (Eq. 3) are *all* needed to account for the frequency dependence of the scattering rate of YBCO$_y$ and Ba122 compounds. Nevertheless, there is a considerable uncertainty with the magnitude of $\Delta_{PG}$ that is as large as $\pm$100 cm$^{-1}$. Irrespective of the uncertainty, the magnitude of this low-energy pseudogap is much smaller than that of high-energy pseudogap which involves the spectral weight transfer over an energy scale of ~ 8000 cm$^{-1}$. Furthermore, the magnitude of $\Delta_{PG}$ in the Ba122 system dramatically exceeds the superconducting energy gap. We will discuss details of the spectral changes related to the high-energy pseudogap in Sec. VI. A.

**D. Infrared pseudogap, antiferromagnetic fluctuations, and antiferromagnetic precursor**

The analysis shown in Fig. 3 reveals a notable correlation between the evolution of the resonant bosonic mode and of the infrared pseudogap in the DOS. This conclusion holds both for YBCO$_{6.65}$ and P-Ba122 systems. In both YBCO$_{6.65}$ and P-Ba122, as the sharp mode in $\alpha^2F(\omega)$ corresponding to the resonant magnetic excitation at AFM wavevectors of the parent compounds becomes weaker, the depth of the infrared pseudogap is suppressed as well. Doping-dependent infrared studies of the cuprates and Ba122 systems reveal the same correlation. Hwang *et al*. reported the results of the electron-boson-coupling analysis of Bi$_2$Sr$_2$CaCu$_2$O$_{8+\delta}$ in wide doping range.[92, 93] In underdoped and OPD compounds, the scattering rate data exhibit a threshold structure due to the formation of the infrared pseudogap. The corresponding $\alpha^2F(\omega)$ spectra show a clear resonance mode associated with magnetic excitations. In the OD compound, the $1/\tau(\omega)$ spectra do not show the threshold structure but are dominated by nearly parallel vertical offset of the entire spectra with the variation in $T$. The same interdependence holds for the Ba122 system.[16] In parent and OPD Ba122 compounds, the sharp mode in $\alpha^2F(\omega)$ spectra becomes weaker with increasing $T$, and the depth of the pseudogap is reduced simultaneously. In OD Ba122, the mode in $\alpha^2F(\omega)$ is very weak and there is no need to invoke a pseudogap in order to explain the behavior of $1/\tau(\omega)$.

We further note that the evolution of the infrared pseudogap across the phase diagram found in our spectra for Ba122 system correlates with that of the AFM instability observed in neutron scattering and nuclear magnetic resonance experiments.[94-96] Both sets of experiments on BaFe$_2$As$_2$ showed the presence of the AFM spin correlations at $T>T_N$. The persistence of the AFM precursor associated with the SDW instability in OPD compounds was demonstrated in nuclear magnetic resonance (NMR) experiments.[95, 96] In OD compounds the AFM precursor is strongly suppressed.[95, 96] In addition, the energy scale of the infrared pseudogap in OPD Ba122 compounds is similar to that of the SDW gap of the parent compound.[16] In undoped Ba122, the pronounced optical transition related to SDW gap formation appears in $\sigma_1(\omega)$ between 500 and 1500 cm$^{-1}$. It is in this frequency region where we observe the infrared pseudogap feature in the conductivity data of OPD compounds: a shallow onset followed by featureless conductivity.

Before closing this section, we discuss an issue on the sharp peak in the $\alpha^2F(\omega)$ spectra adopted from the resonant magnetic excitation in neutron experiments.[83-85] The resonant excitations are significantly enhanced at $T<T_c$ in both YBCO$_y$ and Ba122 systems.[83-85]. For both the systems, the resonant excitations are found to be located at AFM wavevectors of their respective parent phases. While these observations may indicate close relationship between AFM spin fluctuations and superconductivity, we caution that in a conventional excitonic scenario for spin resonance,[97] this effect is linked to a feedback from superconductivity and as such cannot be regarded as an argument for spin-mediated pairing.[98]

## V. RELATION BETWEEN SUPERCONDUCTIVITY AND LOW ENERGY INFRARED PSEUDOGAP

It is instructive to discuss the low-energy infrared pseudogap in the context of the optical fingerprints of superconductivity in both families of high-$T_c$ superconductors. The far-infrared signature of superconductivity (Fig. 4(b)) is the suppression of the optical conductivity in the frequency region where the energy gap forms.[7] The missing spectral weight is condensed into the $\delta$ peak at zero frequency, forming the superconducting condensate. The magnitude of the superconducting gap $2\Delta_{SC}$ is associated with the onset of the optical conductivity for dirty superconductors. In clean superconductors, the absorption structure can be expected at $\omega=4\Delta_{SC}$.[99, 100] Figures 5(a) and 5(b) show that the optical conductivity of YBCO$_{6.50}$ and OPD Co-Ba122 is suppressed below 100 cm$^{-1}$, indicating the superconducting gap opening. We report the magnitudes of superconducting energy gaps in Table I with the values of the SDW gap and the infrared pseudogap for Ba122 compounds.

The onset of superconductivity can be identified in the scattering rate spectra as well. In the frequency region where the optical conductivity is suppressed due to the superconducting gap opening, the scattering rate reveals a distinct suppression.[101] Figures 5(c) and 5(d) show that these expected trends are clearly reproduced in $1/\tau(\omega)$ spectra of both the heavily underdoped YBCO$_{6.50}$ and OPD Co-Ba122. On entering the superconducting state, $1/\tau(\omega)$ is suppressed below 100 cm$^{-1}$ due to the formation of superconducting gap. In a fully gapped superconductor $1/\tau(\omega)$ defined through Eq. (1) is expected to vanish at frequencies below $2\Delta_{SC}$[45] irrespective of impurity scattering (Fig. 4(h)) . We observe that the scattering rate of OPD Co-Ba122 approaches zero near $2\Delta_{SC}$=50 cm$^{-1}$.

Both the scattering rate and the conductivity data reveal that the superconducting gap and the infrared pseudogap features are observed at very different energy scales in YBCO$_{6.50}$ and OPD Co-Ba122. In YBCO$_y$ systems, the difference between $2\Delta_{SC}$ and $\Delta_{PG}$ is largest in heavily underdoped compounds.[55] Other spectroscopic methods capable of probing the energy gaps also show clear separation between the energy scales associated with the infrared pseudogap and superconductivity.[12] Together with the correlation between the infrared pseudogap and magnetism in YBCO$_y$ and Ba122 systems discussed in Sec. IV. D, the scattering rate and the conductivity data suggest that the low-energy infrared pseudogap is unlikely to be related to superconductivity.

## VI. LOW- and HIGH-ENERGY PSEUDOGAPS IN CHARGE AND SPIN EXCITATIONS OF PNICTIDES

An increasing number of experimental papers report on the pseudogap in the iron pnictides. It is instructive to compare both the low-energy infrared pseudogap and high-energy pseudogap in our data with other experimental results in the Fe-SCs. The energy scale of the infrared pseudogap found in our data of the Ba122 compound is about 700 cm$^{-1}$. Infrared studies of the pnictides including our own work show another gap-like feature occurring at much higher energies. The former and the latter are referred to as low- and high-energy pseudogaps of the pnictides, respectively. In this section, we discuss the imprint of the two pseudogaps on the electronic response in the Ba122 systems. We also examine the relationship between the charge and spin pseudogaps.

## A. Charge gap

The term "pseudogap" was used to discuss effects associated with the spectral weight transfer from low to high frequencies with decreasing $T$ in $\sigma_1(\omega)$ of Co-Ba122 compounds in Ref. 102. Spectroscopic features discussed in Ref. 102 are distinct from the low-energy infrared pseudogap found in our data. The energy scale of the spectral weight transfer in Ref. 102 was about 1 eV (~ 8000 cm$^{-1}$); hence the term "high-energy pseudogap" is appropriate.[102] It should be noted that the spectral weight transfer over the energy scale of 1 eV occurs irrespective of the doping level in Co-Ba122[102] and may be a consequence of the multiband electronic structure of these materials.[60] These effects are also registered in our own conductivity data shown in Fig. 6(a).[37] In Fig. 6(b), we plot the spectral weight at low temperature $K(\omega, T)$ normalized to the room temperature value $K(\omega, 295\ K)$. This ratio highlights the energy scale of the spectral weigh transfer. As $T$ decreases, the ratio exceeds 1 at frequencies below about 400 cm$^{-1}$. This is an expected consequence of the narrowing of the Drude-like mode. The ratio falls below 1 at $\omega > 400$ cm$^{-1}$, indicating the depletion of the overall low-energy spectral weight. By 8000 cm$^{-1}$ (~ 1 eV), the ratio returns to the level of 1 signaling that all of the spectral weigh is finally recovered.

Apart from differences in the energy scales, we find other distinctions between the low- and high-energy pseudogaps in Ba122 compounds. Notably, the high-energy pseudogap persists at all dopings.[102] In contrast, the low-energy infrared pseudogap in Fig. 3 shows clear doping dependence. Indeed, the low-energy pseudogap is observed in the parent and OPD compounds but not in the OD compound.[16] These contrasting trends suggest that the high-energy pseudogap reported in Ref. 102 is fundamentally different from the infrared pseudogap in Fig. 3. Furthermore, the spectral weight transfer process (or the high-energy pseudogap following the terminology of Ref. 102) was attributed to the localization of itinerant carriers by Hund's coupling.[37] We note that the magnitude of the Hund's coupling energy in the pnictides is very similar to the energy scale of the spectral weight transfer in $\sigma_1(\omega)$ of Co-Ba122 system.[37]

The presence of the two different low- and high-energy pseudogaps has also been observed in the electron-doped cuprates.[103] Figure 6(c) illustrates the formation of the two pseudogaps in $\sigma_1(\omega)$ of $Nd_{2-x}Ce_xCuO_4$.[103] With decreasing $T$, the conductivity spectra show pseudogap structures at 500 cm$^{-1}$<$\omega$<800 cm$^{-1}$ and 2000 cm$^{-1}$<$\omega$<4000 cm$^{-1}$. To highlight the spectral weight transfer due to the formation of the two pseudogaps, we calculated the $K(\omega, T)/K(\omega, 295\ K)$ ratio for $Nd_{2-x}Ce_xCuO_4$ using the data in Ref. 103. Figure 6(d) shows the result of our analysis. As $T$ decreases, the Drude-like mode narrows leading to $K(\omega, T)/K(\omega, 295\ K)$>1 at $\omega$<10 cm$^{-1}$. We note that the effects of the low-energy pseudogap formation to the spectral weight transfer are minimal. It is the formation of the high-energy pseudogap that leads to the spectral weight transfer to higher energies. The spectral weight ratio falls below 1 at $\omega$>1400 cm$^{-1}$. There exists a minimum in the ratio near 2300 cm$^{-1}$, indicating a turning point of the spectral weight redistribution process. The ratio returns to 1 at about 8000 cm$^{-1}$. The spectral weight analysis shows that all features of the $K(\omega, T)/K(\omega, 295\ K)$ are remarkably similar for $Nd_{2-x}Ce_xCuO_4$ and Ba122 systems.

The presence of the low-energy pseudogap was also reported in an ultrafast pump-probe study of Co-Ba122.[104] The charge gap leads to an increase in the relaxation time of photoexcited quasiparticles. Thus the gap in the electronic density of states is manifested by a tail in the transient electronic response. Stojchevska *et al.* observed a tail in the transient reflectivity data of Co-Ba122 compounds at $T>T_N$, $T_c$, which was attributed to a normal-state pseudogap. The magnitude of the pseudogap was estimated to be ~443 cm$^{-1}$ for OPD Co-Ba122. We note that the magnitude of the pseudogap from the pump-probe experiments is smaller than that in our infrared data. They also observed the two-fold anisotropy in the relaxation dynamics of the crystals showing the pseudogap, which we will discuss in Section VII. B. The presence of the low-energy pseudogap of similar magnitude was also identified in ultrafast experiments of OPD SmFeAsO$_{0.8}$F$_{0.2}$.[105] Importantly, the sign of the reflectance change due to the pseudogap is found to be opposite to that due to the superconducting gap. This difference in the sign of the photo-induced response between the superconducting gap and the low-energy pseudogap also holds for YBCO$_{6.5}$.[106]

A number of infrared studies of hole-doped pnictides, Ba$_{1-y}$K$_y$Fe$_2$As$_2$ (K-Ba122), claimed an observation of the low-energy infrared pseudogap in the conductivity data.[107, 108] Specifically, a gaplike structure in the frequency region between 100 and 200 cm$^{-1}$ was observed in $\sigma_1(\omega)$ of underdoped K-Ba122.[107] The gap-like feature developed at $T \leq 20$ K, which is just above $T_c$. The formation of this feature depletes the spectral weight at $\omega < 110$ cm$^{-1}$ and transfers this weight to the energy region 110 cm$^{-1} < \omega < 250$ cm$^{-1}$. Below $T_c$ the overall spectral weight in the frequency region between 110 and 250 cm$^{-1}$ decreases. From this observation, it was suggested that the pseudogap in the response of the hole-doped K-Ba122 material can be associated with a precursor of superconductivity.[107] The temperature region and the energy scale of the latter pseudogap in $\sigma_1(\omega)$ of hole-doped Ba122 (K-Ba122) is quite different from that of electron-doped Ba122 (Co-Ba122). The hole-doped Ba122 compounds may be more complicated than the electron-doped counterparts because of even stronger propensity to phase separation. It is known that K-Ba122 system is susceptible to inhomogeneity and/or disorder associated with microscopic variation in K ion concentration.[109] We finally note that the suppression in the scattering rate below about 250 cm$^{-1}$ was registered in OPD K-Ba122. The behavior of the scattering rate spectra was analyzed using Eliashberg theory.[61] The Eliashberg analysis accounting for both the pseudogap and charge-boson coupling can identify the pseudogap in hole-doped Ba122 materials.

ARPES studies of Ba122 compounds also reported the presence of a low-energy pseudogap.[110, 111] Shimojima *et al*. reported on the laser ARPES data of P-Ba122 system.[110] They observed that the pseudogap: a depletion of the spectral weight near the Fermi energy, develops well above the AFM transition in the parent and underdoped compounds and persists above the nonmagnetic superconducting dome. In the overdoped regime, the pseudogap vanished from the data. The magnitudes of the pseudogap in underdoped $BaFe_2(As_{0.93}P_{0.07})_2$ and OPD $BaFe_2(As_{0.7}P_{0.3})_2$ were estimated to be about 970 cm$^{-1}$ and 480 cm$^{-1}$, respectively. The latter value is comparable to that extracted from our infrared data of OPD P-Ba122 with the same composition. The ARPES data showed that the magnitude of the superconducting gap (~ 80 cm$^{-1}$) is much smaller than that of the pseudogap. The evolution of the pseudogap with temperature and doping registered in the ARPES data correlates with that of the electronic nematicity in the magnetic torque experiments.[112] suggesting the intimate relationship between the pseudogap and the nematic fluctuations. The pseudogap behavior was also observed in the photoemission spectra of $Ba_{0.75}K_{0.25}Fe_2As_2$ ($T_c$=26 K).[111] The magnitude of the pseudogap was about 290 cm$^{-1}$, which is larger than that of superconducting gap by a factor of 2 – 4. The authors showed that the pseudogap occurs mainly on the Fermi surface sheets that are connected by the AFM wave vector of the Ba122 compound. From these observations, they suggested that the pseudogap may be attributed to the AFM fluctuations.

**B. Spin gap**

NMR is a useful tool to probe (spin) pseudogap in unconventional superconductors. This technique was the first to observe the pseudogap in the spin channel of underdoped YBCO$_y$.[9, 113] Manifestations of the pseudogap can be found in the Knight shift or spin-lattice relaxation rate data. In a Fermi-liquid picture, the Knight shift is proportional to the DOS at the Fermi level. The spin-lattice relaxation rate is governed by electronic excitations and can reveal the indirect signatures of the electronic pseudogap. NMR studies of the cuprates identified a decrease in the two quantities below $T^*$ reflecting the decrease in the electronic DOS.[9] The development of the electronic pseudogap first found by infrared spectroscopy shows close correlation with that of the spin pseudogap in NMR data.[64, 114]

NMR studies of the pnictides have provided information on the pseudogap state. The effects seen in NMR data that are attributed to the pseudogap of Co-Ba122 are different from what is observed in the cuprates. The Knight shift does not show the suppression at a particular $T$. Instead, as $T$ decreases, the Knight shift is gradually suppressed with a minor slope change at a certain $T$ hindering the determination of the pseudogap temperature.[50, 95] The magnitude of the pseudogap was estimated by fitting the $T$ dependence of the Knight shift data with the empirical activation formula $A + B\exp(-\Delta_{NMR}/k_B T)$.[50, 95] The value of $\Delta_{NMR}$ ranges from 480 cm$^{-1}$ ($x$=0.0) to 300 cm$^{-1}$ ($x$=0.26). The magnitude of $\Delta_{NMR}$ is summarized along with those of $\Delta_{PG}$, $\Delta_{SDW}$ (spin-density-wave gap), and $2\Delta_{SC}$ in Table I. It appears that the spin pseudogap inferred from NMR data does not exactly correspond to the low-energy pseudogap in our infrared data. Furthermore, whereas the infrared pseudogap disappears in OD compound ($x$=0.25),[16] the spin pseudogap is still present in Co-Ba122 with $x$=0.26.[95]

The spin-lattice relaxation rate of the parent Ba122 compound does not display the depression familiar from the studies of the underdoped cuprates. Instead the spin-lattice relaxation rate is drastically enhanced with decreasing $T$.[95] The enhancement was interpreted as a signature of strong AFM precursor related to the SDW instability of the parent material. The AFM precursor were found to persist in OPD compound but disappear in the OD compound ($x \geq 0.15$).[95] It is interesting to note that the evolution of the infrared pseudogap in our data correlates with the trends seen through the analysis of AFM precursor but not with spin-pseudogap in the NMR data.

Unlike Co-Ba122, the spin-lattice relaxation rate data for Ca(Fe,Co)$_2$As$_2$ do not show strong enhancement at low $T$.[115] Instead these latter data exhibit the suppression at $T^*>T_N$, similar to the pseudogap feature in the spin-lattice relaxation rate data of the cuprates. Baek *et al.* argued that the first-order character of the SDW transition in Ca(Fe,Co)$_2$As$_2$ may inhibit strong AFM precursor which are responsible for the enhancement of the spin-lattice relaxation. As a result, the suppression in the spin-lattice relaxation can be apparent.[115] The difference in the spin-lattice relaxation rate in Co-Ba122 and Ca(Fe,Co)$_2$As$_2$ suggests that the comparison between the two systems might not be straightforward.

### C. Interlayer response

Manifestations of the pseudogap can be identified in *c*-axis transport and infrared data. In the cuprates, as the pseudogap develops below $T^*$, the *c*-axis resistivity $\rho_c$ is enhanced above linear trend and often shows a "semiconducting" behavior.[116] Tanatar *et al.* reported $\rho_c$ data of Co-Ba122 over broad doping range.[52] They observed two anomalies in the $T$ dependence of $\rho_c$; a crossover from metallic to nonmetallic behavior at $T_{CG}$ and another crossover from nonmetallic to metallic behavior at $T^{**}$ with further decrease of $T$. The temperature scale $T_{CG}$ was attributed to the formation of a charge gap at the Fermi surface (electronic pseudogap). The values of $T_{CG}$ in both parent and OPD compounds were expected to be above 300 K, which appears to be consistent with the infrared data in Fig. 3. Upon doping, $T_{CG}$ gradually decreases and vanishes at $x\approx0.3$; the open triangles in Fig. 7(b) represent $T_{CG}$. It should be pointed that the *c*-axis transport data suggest the presence of the charge gap at $x=0.25$ where our *ab*-plane infrared data show no pseudogap.[16] The other temperature scale $T^{**}$ (open squares in Fig. 7(b)) coincides with the temperature at which the Knight shift shows a slope change and the spin-lattice relaxation rate increases. This correspondence suggests that magnetic correlations play an important role in the anomalies in the resistivity, Knight shift, and spin-lattice relaxation data.

In the cuprates, the pseudogap formation was most clearly identified in the $c$-axis charge dynamics.[64, 114, 117, 118] Above $T^*$, the $c$-axis optical conductivity $\sigma_{1c}(\omega)$ of YBCO$_{6.70}$ is flat with frequency. As $T$ decreases across $T^*$, the $c$-axis conductivity at low frequencies is gradually suppressed revealing a gaplike structure.[64] We stress that very similar behavior was observed in $\sigma_{1c}(\omega)$ of OPD Co-Ba122.[119] These latter experiments revealed a continuous depression in flat and featureless $\sigma_{1c}(\omega)$ with decreasing $T$.[119] The distinct characteristic of the $c$-axis charge dynamics of OPD Co-Ba122 discriminating the latter from underdoped cuprates is the coherent Drude-like response. The coherent Drude-like peak coexists with the pseudogap-like depression at the lowest $T$. The coherent response can be linked to the metallic behavior of the $c$-axis resistivity at low $T$.[52]

# VII. ON THE NATURE OF THE PSEUDOGAP PHASE IN CUPRATES AND PNICTIDES

## A. Pseudogap and Superconductivity

The remarkable similarities of the electrodynamics revealed in our infrared data for the prototypical cuprate and Fe-based materials (Fig. 3) establish the relevance of the pseudogap phase to the phenomenology of the high-$T_c$ pnictides. Further, these results provide clues on the origin of the pseudogap state in both systems. The evolution of the electrodynamics across $T_c$ shows that the energy scale of the low-energy infrared pseudogap is well separated from that of the superconducting gap in YBCO$_y$ and Ba122 systems. This finding suggests no direct connection between the pseudogap and high-$T_c$ superconductivity in either the cuprates or the Fe-SCs. We also remark that despite the difference in the symmetry of superconducting gap of YBCO$_y$ and Ba122 systems the similar pseudogap feature is observed. The analysis reported in Section IV. C demonstrated the intimate connection between the infrared pseudogap and magnetism of the parent phases: yet another similarity of the cuprates and the Fe-SCs. In the case of the Fe-SCs, we found that the energy scale of the infrared pseudogap is comparable to that of the spin-density-wave gap in the parent phase. The totality of our infrared data suggests that antiferromagnetism is a likely cause of the pseudogap in the two classes of high-$T_c$ superconductors.

### B. Nematic Fluctuations

The nematic correlations leading to the anisotropic response of CuO$_2$ planes are considered as one of the potential candidates for the origin of the low-energy pseudogap in the cuprates.[30, 120-123] Daou *et al.* observed a large in-plane anisotropy of the Nernst effect in YBCO$_y$ that sets in precisely at $T^*$.[120] A scanning tunneling microscopy study of underdoped Bi$_2$Sr$_2$CaCu$_2$O$_{8+\delta}$ revealed intra-unit-cell electronic nematicity which was attributed to the weak magnetic states at the O sites that break 90° rotational ($C_4$) symmetry within the CuO$_2$ unit cell.[121] Very recently, from a spin-fermion model analysis, several groups suggested the relevance of a quadrupole-density wave order to the pseudogap in the cuprates.[30] In solving the spin-fermion model for the quantum antiferromagnet-normal-metal transition in a system of two-dimensional itinerant electrons, a state with a gap over some part of the Fermi surface without long-range order was found at finite temperatures. It was suggested that this state may be a superposition of *d*-wave superconductivity and a quadrupole-density wave. The quadrupole-density wave is a pure charge order involving charge modulation on the four O atoms surrounding a Cu atom. In real space, the quadrupole-density wave corresponds to a chequerboard pattern, which is similar to the charge modulation associated with the intra-unit-cell electronic nematicity observed in the scanning tunneling microscopy data of underdoped $Bi_2Sr_2CaCu_2O_{8+\delta}$.[121]

Recent ARPES, Kerr effects, and time-resolved reflectivity experiments on nearly OPD $Pb_{0.55}Bi_{1.5}Sr_{1.6}La_{0.4}CuO_{6+\delta}$ uncovered signatures of the broken symmetry in the pseudogap state, suggesting that the pseudogap temperature $T^*$ is associated with a phase transition.[122] Inelastic neutron scattering experiments on $YBCO_{6.45}$ also suggested that the onset of the pseudogap is related to a phase transition that involves electronic nematic phase.[124] Hinkov *et al*. observed the spontaneous onset of a one-dimensional incommensurate modulation of the spin system upon cooling below 150 K. The evolution of the modulation with temperature and doping parallels that of the in-plane resistivity anisotropy.

The nematicity may also be relevant to the pseudogap of the Fe-SCs. Transport and infrared spectroscopy studies of detwinned Co-Ba122 have identified anisotropy of the in-plane electronic response[125-128] in the parameter space where our infrared data detect the pseudogap. The parent compound of the pnictides shows stripe-type AFM order[14, 15] and the spin nematicity is observed in the paramagnetic state of the parent and doped compounds.[129] Our infrared data indeed show that the pseudogap is related to the SDW instability toward the stripe-type spin ordering within Fe planes.

Electronic anisotropy has been suggested in recent time-domain optical spectroscopy study of Co-Ba122.[104] In this work, linearly polarized light was used for pump-probe experiments. The transient response of photoexcited quasiparticles shows a two-fold rotational anisotropy that persists up to about 200 K in the parent, underdoped, and OPD crystals. The transient reflectivity data also displayed the bottleneck in the relaxation of photoexcited electrons at $T>T_N$, indicating the presence of the low-energy pseudogap. These results are indicative of a connection between the two-fold electronic anisotropy and the pseudogap. However, on a closer inspection one notices differences between the results of the two spectroscopies. For example, in OD Co-Ba122 with $x$=0.11, the relaxation bottleneck is present but the electronic anisotropy is absent.

Spontaneous $C_4$ symmetry breaking associated with the nematicity has been clearly demonstrated in recent magnetic torque measurements of P-Ba122.[112] Kasahara *et al.* observed the appearance of two-fold oscillation of magnetic torque above structural/magnetic transition temperatures and its persistence to nonmagnetic superconducting regime. The temperatures below which the two-fold oscillations are observed are shown with solid circles in the phase diagram of the pnictides (Fig. 7(b)). A transport study of P-Ba122 showed that the application of mechanical stress enhances the nematic susceptibility.[130] Ultrasound spectroscopic studies have also indicated that the nematic fluctuations determined from the elastic properties of Co-Ba122 are prominent over a large portion of the phase diagram of the pnictides.[131] Scanning tunneling microscopy studies of underdoped $Ca(Fe_{1-x}Co_x)_2As_2$ found a unidirectional electronic nanostructure with a dispersive *b*-axis quasiparticle interference modulations associated with nematic 180° rotational ($C_2$) symmetry in the low-temperature AFM state.[125] Spatially-averaged DOS exhibit a V-shaped pseudogap spectrum in this state. Scanning tunneling microscopy experiments in the paramagnetic state may provide direct information on the relation between the pseudogap and nematicity.

The origin of the nematicity in the pnictides has been discussed in the context of the precursors of either spin or orbital orderings. The magnetic scenario has been discussed in detail by Fernandes *et al.*[132] The spin nematicity was indeed observed in the neutron scattering measurements of Co-Ba122.[129] ARPES study of strained Co-Ba122 demonstrated degeneracy lifting between $d_{yz}$ and $d_{zx}$ orbital states which can induce the in-plane anisotropy, consistent with the results from other probes.[133] Point-contact spectroscopy data of 122 pnictides revealed the enhancement of zero-bias conductance in parent and Co-underdoped Ba122 below $T=T_{pc}$, which are shown with solid diamond in Fig. 7(b).[134]

Lee and Phillips found that the enhancement of zero-bias conductance in the point contact spectroscopy experiments[134, 135] can be explained in terms of the precursor of the orbital ordering, i.e., orbital fluctuations.[136] The enhancement of zero-bias conductance were only observed at the dopings and temperatures[134] where the in-plane resistivity anisotropy exists.[125] On the other hand, Valenzuela *et al.* suggested that the orbital ordering favors the resistivity anisotropy opposite to what is found experimentally.[137] It is worth pointing out that the in-plane resistivity anisotropy is anomalous: the resistivity along the longer axis (AFM direction) is smaller than that along the shorter axis (FM direction). Investigations of the polarization dependence of the infrared pseudogap may elucidate the relation between the pseudogap and spin/orbital fluctuations associated with the in-plane anisotropy.[127] We finally note that the nematicity can invoke profound impact on superconducting pairing, such as an increase in $T_c$ or a change in the symmetry of the superconducting gap.[138]

## VIII. SUMMARY AND OUTLOOK

We analyzed the optical spectra for the two prototypical families of the cuprate and iron pnictide superconductors: YBCO$_y$ and Ba122 systems, with an emphasis on the pseudogap characteristics. The optical data reveals the presence of the low- and high-energy pseudogaps in the Ba122 systems. The formation of the high-energy pseudogap involves spectral weight transfer from low to high frequencies with a typical energy scale is about 1 eV. The similar high-energy pseudogap behavior was observed in an electron-doped cuprate Nd$_{2-x}$Ce$_x$CuO$_4$.[103] The infrared signatures of the low-energy pseudogap in YBCO$_y$ and Ba122 compounds share striking similarities. The low-energy infrared pseudogap produces a shallow dip at about 500 cm$^{-1}$ in otherwise nearly featureless mid-infrared conductivity. The manifestation of the infrared pseudogap is more evident in the scattering rate data. The formation of the infrared pseudogap leads to the suppression in the scattering rate below characteristic frequencies revealing a threshold structure in $1/\tau(\omega)$ spectra. Our estimate of $\Delta_{PG}$ from the scattering rate is about 700 cm$^{-1}$ for both the compounds. The magnitude of the infrared pseudogap turns out to be much larger than $2\Delta_{SC}$. We performed the theoretical analysis of the scattering rate within a model accounting for the electron-boson coupling and the electronic pseudogap in DOS. The bosonic spectral function used in the analysis of YBCO$_y$ and Ba122 systems possesses two components: a magnetic resonance mode and broad background. The model analysis of the scattering rate reveals close interdependence between the pseudogap and electron-boson coupling; the development of the pseudogap is correlated with the strengthening of the magnetic resonance mode. The low-energy pseudogap in the Ba122 family of materials is revealed in photoemission[110, 111] and pump-probe[104] experiments. The persistence of the gaplike response above $T_c$ was uncovered by tunneling experiments for NaFeAs.[139] Even though numerous experimental studies of the pnictides hint to the sightings of the pseudogap additional experiments are needed to establish the universality of this phenomenon in the Fe-SCs.

## ACKNOWLEDGEMENTS

This work was supported by National Science Foundation (NSF 1005493) and AFOSR. SJM is supported by Basic Science Research Program through the National Research Foundation of Korea funded by the Ministry of Science, ICT & Future Planning (2012R1A1A1013274) and TJ Park Science Fellowship of POSCO TJ Park Foundation. YSL is supported by a National Research Foundation of Korea (NRF) grant funded by the Korea government (MOE) (No. 2013R1A1A2012281). AVC is supported by DOE Grant No. DE-FG02-ER46900. Work at Ames Laboratory (PCC, SLB, MT, RP, AT) was supported by the U.S. Department of Energy, Office of Science, Basic Energy Sciences, Materials Sciences and Engineering Division. Ames Laboratory is operated for the U.S. Department of Energy by Iowa State University under contract DE-AC02-07CH11358. AS acknowledges support by the Department of Energy, Basic Energy Sciences, Materials Sciences and Engineering Division.

**Figures**

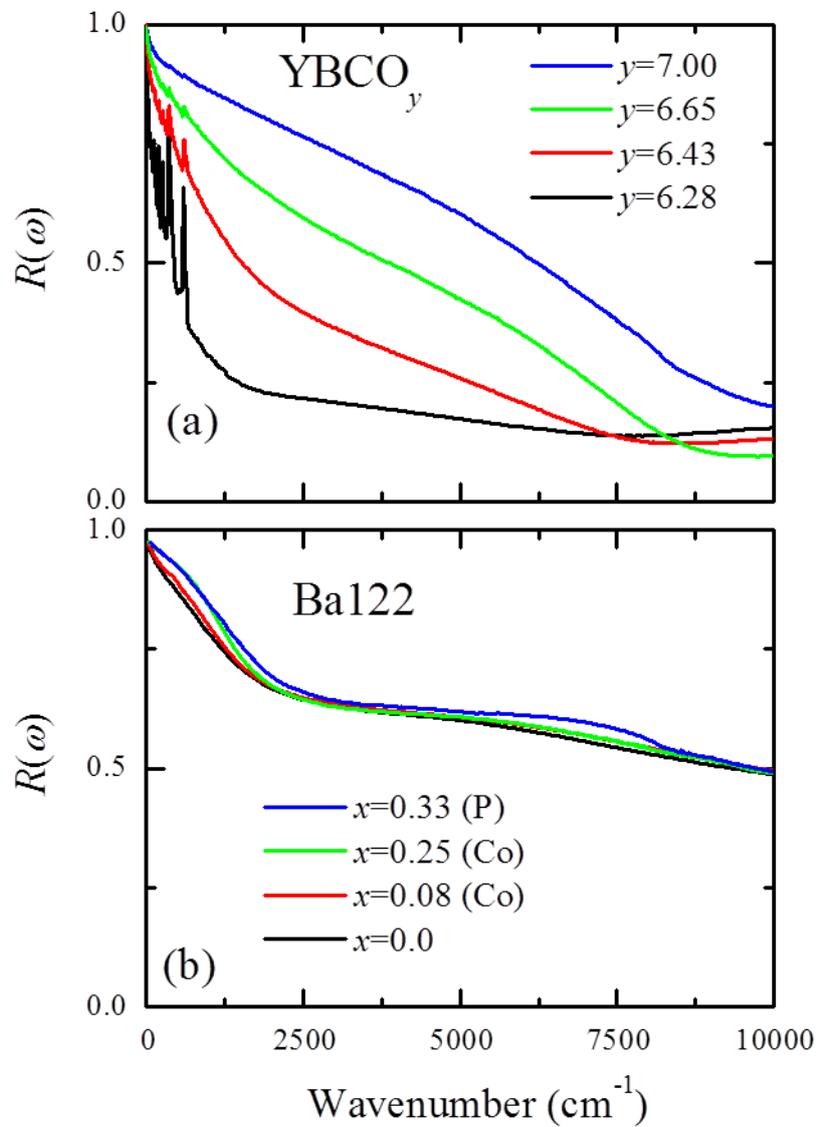

FIG. 1 (color online). Room-temperature in-plane reflectance $R(\omega)$ spectra of YBCO$_y$ (Ref. 55) and Co-/P-doped Ba122.

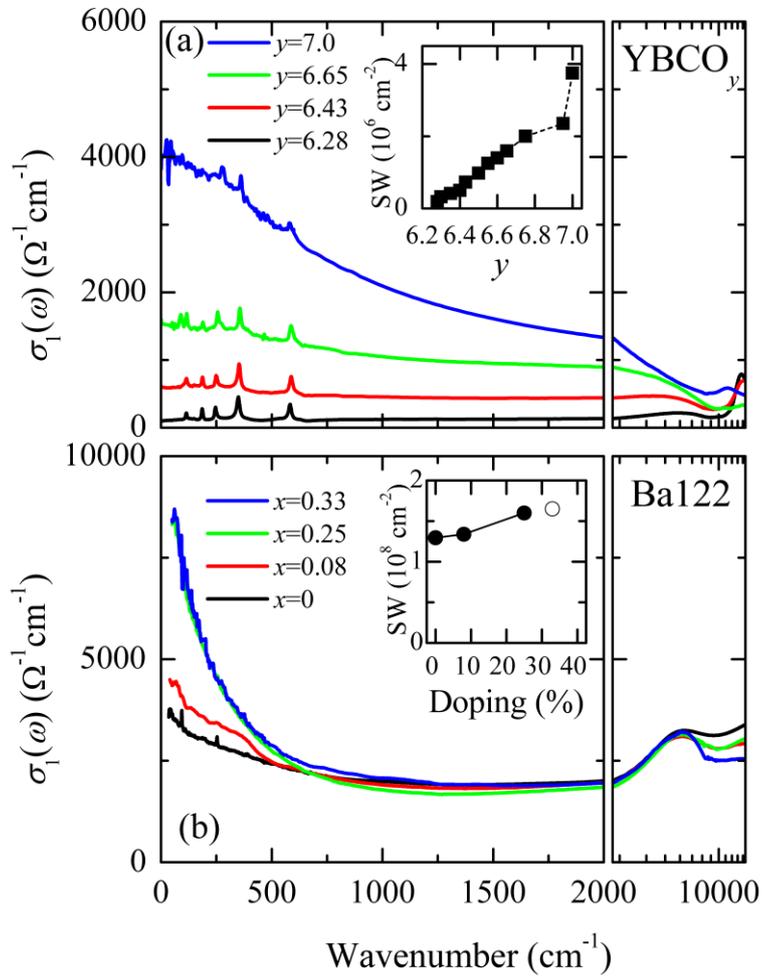

FIG. 2 (color online). Doping dependence of the real part of the in-plane optical conductivity $\sigma_1(\omega)$ spectra of (a) YBCO$_y$ (Ref. 55) and (b) Ba122 systems. Insets of (a) and (b) show the spectral weight obtained from the integration of $\sigma_1(\omega)$ of YBCO$_y$ up to 8000 cm$^{-1}$ and that of Ba122 up to 1500 cm$^{-1}$, respectively.

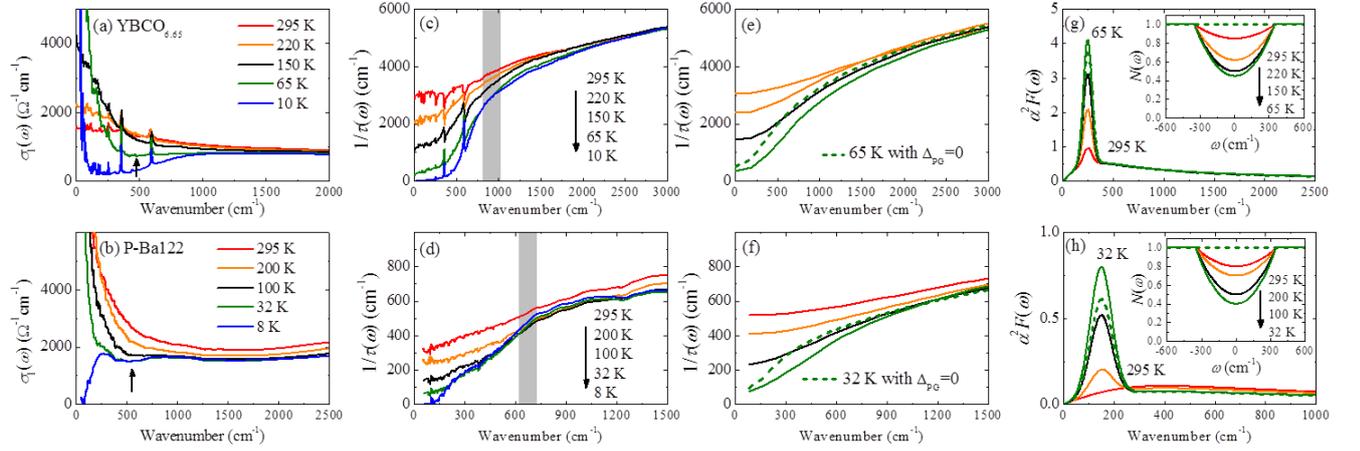

FIG. 3 (color online). In-plane electrodynamics of YBCO$_y$ and Ba122 superconductors. (a)-(b) $T$-dependent optical conductivity. (c)-(d) Scattering rate $1/\tau(\omega)$. (e)-(f) Calculated scattering rate. (g)-(h) Bosonic spectral function $\alpha^2F(\omega)$ employed in the electron-boson-coupling analysis. Insets: normalized density of states. Top panels: YBCO$_{6.65}$. Bottom panels: P-Ba112.

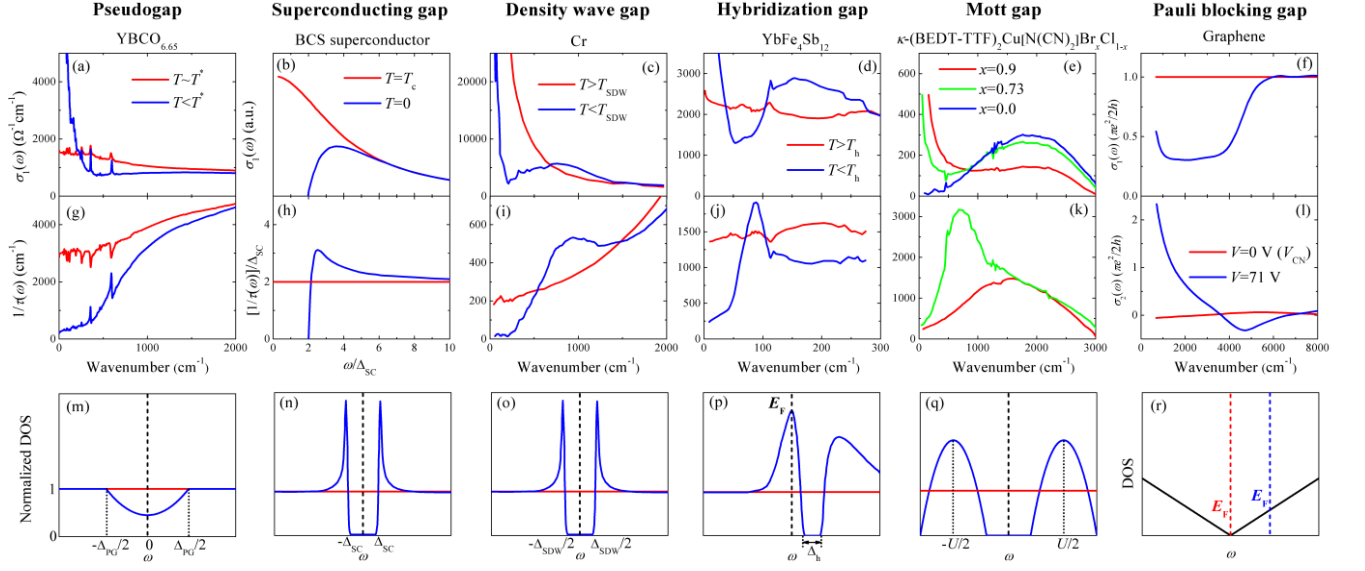

FIG. 4 (color online). Top panels: real part of optical conductivity of (a) YBCO$_{6.65}$, (b) BCS superconductor, (c) Cr, (d) YbFe$_4$Sb$_{12}$, (e) $\kappa$-(BEDT-TTF)$_2$Cu[N(CN)$_2$]Br$_{1-x}$Cl$_x$, and (f) graphene. $T^*$: pseudogap temperature, $T_c$: superconducting transition temperature, $T_{SDW}$: spin-density-wave transition temperature, $T_h$: hybridization gap formation temperature, $V$: applied voltage. Middle panel: scattering rate (g)-(k) and the imaginary part of optical conductivity for graphene (l), Bottom panel: (m)-(q) normalized density of states, (r) density of states. $\Delta_{PG}$: pseudogap, $2\Delta_{SC}$: superconducting gap, $\Delta_{DW}$: density-wave gap, $\Delta_h$: hybridization gap, $U$: on-site Coulomb interaction, $E_F$: Fermi level. Thick dashed line indicates the Fermi level, which is set to zero in (m)-(q). The DOS plot in (o) is for an ideal SDW case where the full gap is developed. In Cr, some bands are not impacted by SDW order and there remains finite DOS at the Fermi level in the SDW state. Cr (Ref. 45), YbFe$_4$Sb$_{12}$ (Ref. 68), $\kappa$-(BEDT-TTF)$_2$Cu[N(CN)$_2$]Br$_{1-x}$Cl (Refs. 69, 70), graphene (Ref. 74).

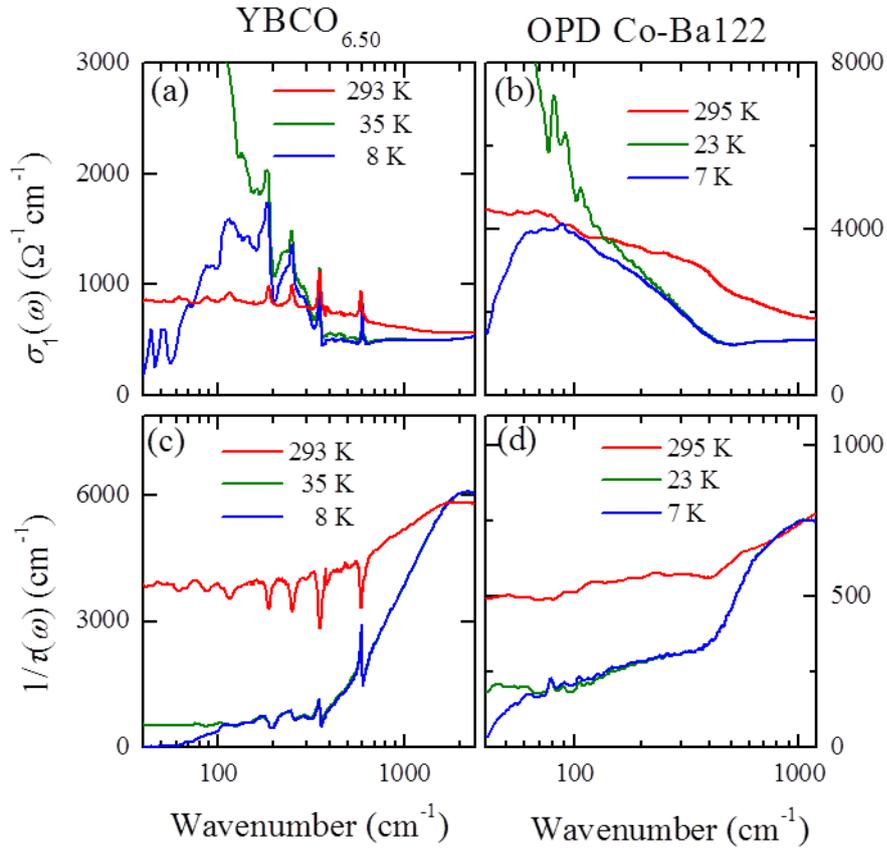

FIG. 5 (color online). Optical conductivity (top) and scattering rate (bottom) for the underdoped YBCO$_{6.50}$ (left column) and OPD Co-Ba112 (right column).

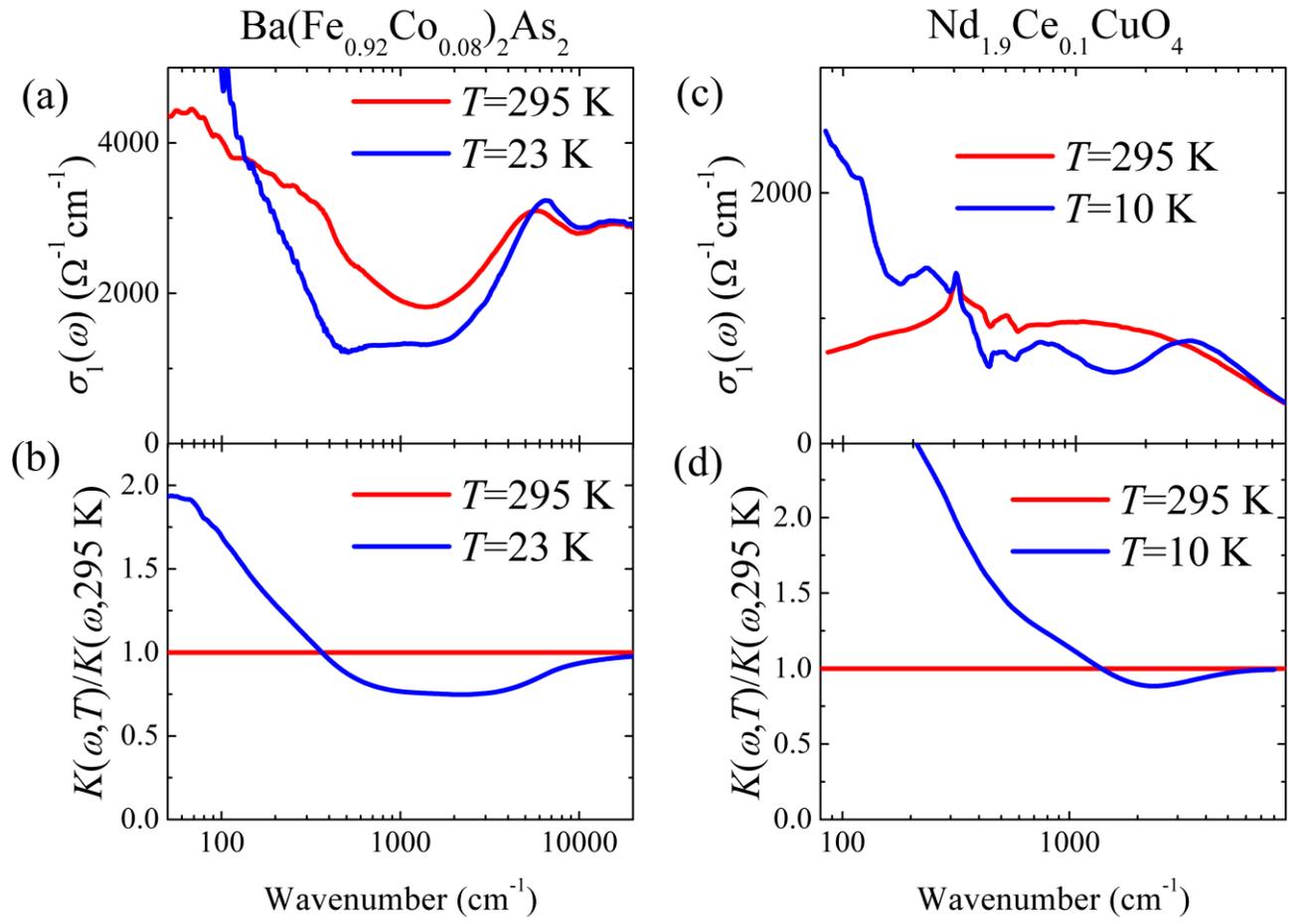

FIG. 6 (color online). Optical conductivity spectra of (a) Ba(Fe$_{0.92}$Co$_{0.08}$)$_2$As$_2$ and (c) Nd$_{1.9}$Ce$_{0.1}$CuO$_4$ (Ref. 103). Ratio of the spectral weight $K(\omega, T)/K(\omega, 295\,\text{K})$ of (c) Ba(Fe$_{0.92}$Co$_{0.08}$)$_2$As$_2$ and (d) Nd$_{1.9}$Ce$_{0.1}$CuO$_4$.

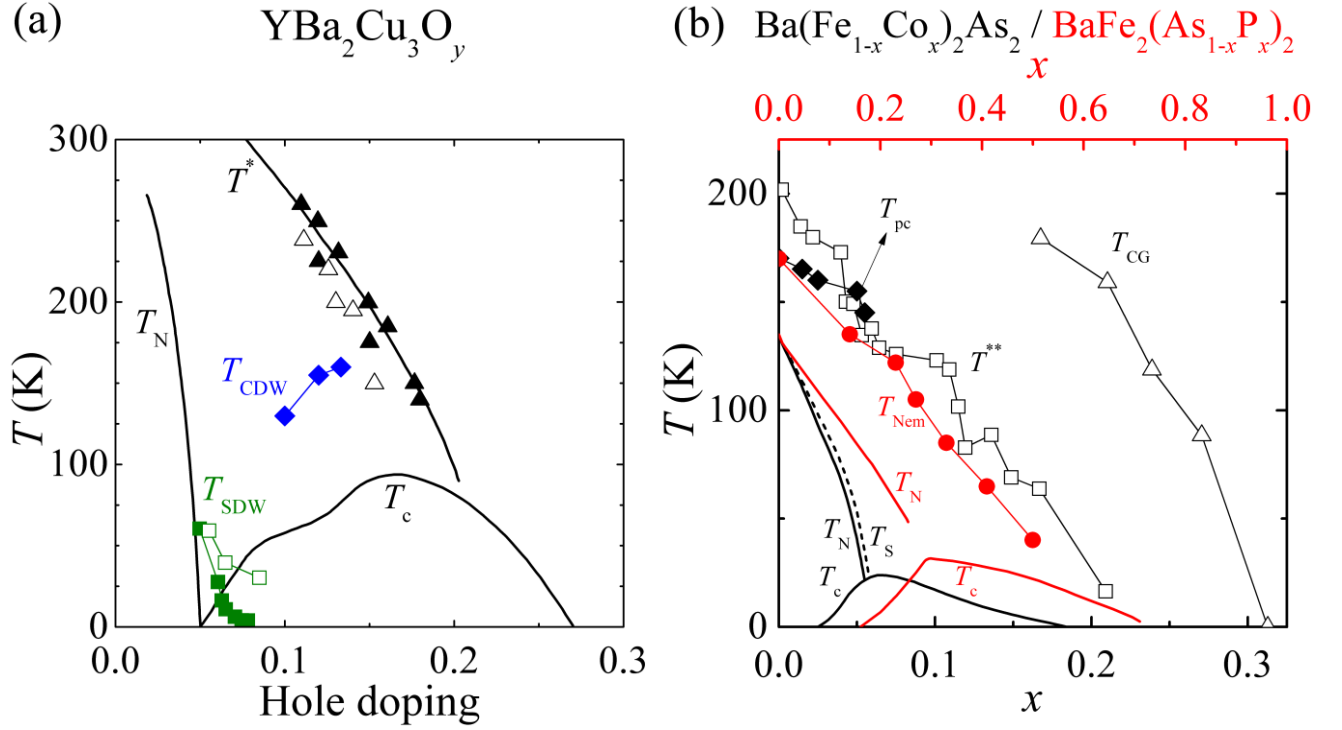

FIG. 7 (color online). (a) Phase diagram of YBCO$_y$ (Refs. 25, 140). $T^*$: the pseudogap onset temperature determined from the in-plane resistivity (open triangles, Ref. 141) and Nernst experiments (solid triangles, Ref. 142). Solid triangles are $T_{SDW}$: the onset temperature for spin-density-wave order measured by neutron scattering (open squares, Ref. 143) and muon spin rotation experiments (solid squares, Ref. 144). $T_{CDW}$: the onset temperature of the charge-density-wave order measured by x-ray scattering experiments (solid diamonds, Ref. 22). (b) Phase diagram of Co-Ba122 (Refs. 51, 145) and P-Ba122 (Refs. 48, 112). The lower and upper horizontal axes are for P-Ba122 and Co-Ba122, respectively. The boundary of AFM and superconducting states are indicated by thick black lines for Co-Ba122 and thick red lines for P-Ba122. The structural transition temperature $T_S$ of Co-Ba122 is represented by black dashed line. $T^{**}$ (open squares): a crossover temperature at which the behavior of the c-axis resistivity of Co-Ba122 changes from nonmetallic to metallic with decreasing $T$ (Ref. 52). $T_{CG}$ (open triangles): a crossover temperature at which the behavior of the c-axis resistivity of Co-Ba122 changes from metallic to nonmetallic with decreasing $T$ (Ref. 52). $T_{pc}$: a temperature below which the point-contact experiments of Co-Ba122 show conductance enhancement (Refs. 134, 135). $T_{\text{Nem}}$ (solid circles): the temperature below which the two-fold torque component appears in the magnetic torque experiments of P-Ba122 (Ref. 112). In both panels, $T_{\text{N}}$ represents AFM transition temperature and $T_{\text{c}}$ stands for superconducting transition temperature.

| IR | Ba(Fe$_{1-x}$Co$_x$)$_2$As$_2$ | | | BaFe$_2$(As$_{1-x}$P$_x$)$_2$ |
|---|---|---|---|---|
| Co or P concentration | 0 | 0.08 | | 0.33 |
| $\Delta_{PG}$ (cm$^{-1}$) | 700 | 700 | | 700 |
| $\Delta_{SDW}$ (cm$^{-1}$) | 336, 656 [a] | | | |
| $2\Delta_{SC}$ (cm$^{-1}$) | | 50 | | 80 |

| Pump-probe | Ba(Fe$_{1-x}$Co$_x$)$_2$As$_2$ [b] | | | SmFeAsO$_{0.8}$F$_{0.2}$ [c] |
|---|---|---|---|---|
| Co concentration | 0.051 | 0.07 | 0.11 | |
| $\Delta_{PG}$ (cm$^{-1}$) | 538 | 443 | 410 | 492 |

| NMR | Ba(Fe$_{1-x}$Co$_x$)$_2$As$_2$ [d] | | | | |
|---|---|---|---|---|---|
| Co concentration | 0 | 0.04 | 0.08 | 0.105 | 0.26 |
| $\Delta_{NMR}$ (cm$^{-1}$) | 480 | 380 | 350 | 330 | 300 |

TABLE I. Magnitudes of the pseudogap extracted from the analysis of the infrared spectroscopy (top), pump-probe spectroscopy (middle), and nuclear magnetic resonance experiments (bottom). In the top panel, the values of the spin-density-wave gap and superconducting gap are also presented for comparison. a: Ref. 37, b: Ref. 104, c: Ref. 105, d: Refs. 50, 95.